\documentclass[journal,10pt]{IEEEtran}
\usepackage{amsfonts}
\IEEEoverridecommandlockouts

\ifCLASSINFOpdf
\else
\fi
\usepackage{epsfig}
\usepackage{graphicx}
\usepackage{psfig}
\usepackage{epsf}
\usepackage[cmex10]{amsmath}
\usepackage{booktabs}
\usepackage{fancyhdr}
\usepackage{stfloats}
\usepackage{color}
\usepackage{multirow}
\usepackage{array}
\usepackage{caption2}
\newcommand{\tabincell}[2]{\begin{tabular}{@{}#1@{}}#2\end{tabular}}

\begin{document}
\title{Performance Analysis on Visible Light Communications With Multi-Eavesdroppers and Practical Amplitude Constraint} %Random Location%\author{\IEEEauthorblockN{Author 1, Author 2, Author 3, Author 4 and Author 5}
\author{\IEEEauthorblockN{Xiaodong Liu, \emph{Student Member, IEEE}, Yuhao Wang, \emph{Senior member, IEEE}, Fuhui Zhou, \emph{Member, IEEE}, \\  Zhenyu Deng, and Rose Qingyang Hu, \emph{Senior Member, IEEE}}
\thanks{Manuscript received August 20, 2019; revised September 18, 2019; accepted October 11, 2019. This work was supported in part by the National Natural Science Foundation of China (61661028 and 61701214) and the National Key Research and Development Project (2018YFB1404303), in part by the Intel Corporation, and in part by the US National Science Foundation (EARS1547312), in part by the Excellent Youth Foundation of Jiangxi Province (2018ACB21012) and the Postdoctoral Science Foundation of Jiangxi Province (2017KY04). Corresponding authors: Fuhui Zhou; Yuhao Wang. }
\thanks{X. Liu is with the School of Electronic Information, Wuhan University, Wuhan, 430072, China. (e-mail: xiaodongliu@whu.edu.cn).
Y. Wang, F. Zhou, and Z. Deng are with the School of Information Engineering, Nanchang University, Nanchang, 330031, China. (e-mail: wangyuhao@ncu.edu.cn, zhoufuhui@ieee.org, sky312312@126.com).
F. Zhou and R. Q. Hu are with the Department of Electrical and Computer Engineering at Utah State University, Logan, UT 84322 USA. (e-mail: zhoufuhui@ieee.org, rose.hu@usu.edu).}
}
\maketitle
\begin{abstract}
In this paper the secure performance for the visible light communication (VLC) system with multiple eavesdroppers is studied.  By considering the practical amplitude constraint instead of an average power constraint in the VLC system, the closed-form expressions for the upper and the lower bounds of the secrecy outage probability and the average secrecy capacity are derived. Since the locations for both legitimate receiver and eavesdroppers are unknown as well as random, the stochastic geometry method is introduced for problem formulation. Moreover,  accurate expressions with a low computational complexity are obtained with the rigorous derivation. Simulation results verify the correctness of our theoretical analysis.
\end{abstract}
%\vbox{}
\IEEEpeerreviewmaketitle
\section{Introduction}
\IEEEPARstart{V}{ISIBLE} light communication (VLC)  has received increasing attentions due to the abundant license-free spectrum resources and the effective reuse of frequency and space.  It is a promising technology to tackle the spectrum scarcity \cite{VLC_intr_2}, \cite{AMC} and to support the fifth generation and beyond wireless communication systems \cite{VLC_book}. Moreover, different from the traditional radio frequency (RF) channel, VLC is mainly implemented by light-of-sight (LoS) as visible light is blocked by the opaque materials, which is beneficial for secure transmissions. Nevertheless, VLC  faces the risk of eavesdropping due to the broadcast nature \cite{VLC_security_nature}, \cite{SOP_GAO}. To address this issue, two main schemes can be introduced. One hand, upper-layer encryption \cite{VLC_security_nature} is a common  secure scheme, but it is often challenged by high computational power and cloud computing.  On the other hand, physical layer security (PLS), as a new and effective secure scheme, can achieve information secure transmission based on the randomness of channels \cite{CC_MS}, \cite{PLS_channel_2}. % by using the characteristic of wireless channels 

Although many existing works have analyzed the secure performance of PLS techniques in the traditional RF based wireless communication systems, these analyses cannot be directly applied to the VLC systems since the VLC channels are subject to the amplitude constraint \cite{AMC}, \cite{CC_MS}. As such, generally speaking it is difficult to obtain an analytical expression for the VLC  channel capacity \cite{CC_AM}. Fortunately, there have been several closed-form expressions derived the upper and the lower bounds for the channel capacity with the amplitude constraint in the VLC systems \cite{CC_MS}, \cite{CC_AM} and \cite{CC_OWC}. %\cite{PLS_channel_1}, %In order to address this challenge,

Based on the lower bound of the channel capacity derived in \cite{CC_OWC}, the authors in \cite{3D_YL} investigated the PLS in a 3-D multiuser VLC systems with and without the access points (AP) cooperation. Meanwhile, in order to enhance the security, a scheme that builds a disk-shaped secrecy protection zone around the AP was proposed. Moreover, in \cite{CC_AM}, the analytically tractable expressions for the lower and the upper channel capacities were derived, and the achievable secrecy rates of multiple-input-single-output (MISO) VLC  systems were analyzed. A robust beamforming scheme for maximizing the worst-case secrecy rate was proposed. Furthermore, in order to maximize the secrecy rate or minimize the total power consumption, the optimal and robust secure beamforming schemes were proposed in MISO VLC systems under both perfect and imperfect channel state information (CSI) of eavesdropper \cite{CC_MS}.

Note that the secure analysis and robust beamforming in \cite{CC_MS} and \cite{CC_AM} were established in the case that the CSI or the locations of the eavesdroppers are known. However, the locations of the eavesdroppers are usually unknown in practice \cite{SOP_LED}. In this case, the authors have analyzed the secure performance and proposed a new MISO beamforming scheme based on the eavesdropper intensity. Moreover, the closed-form expressions for the bound of the secrecy outage probability (SOP) were derived under the light emitting diode (LED) selection scheme. Furthermore, by ignoring the peak power constraint, the generalized closed-form expressions of the SOP and the average secrecy capacity (ASC) were derived in the VLC system with multiple random distributed eavesdroppers \cite{SOP_GAO}. %in \cite{SOP_LED}

Instead of considering the over-simplified signal model in [5], the secure performance is analyzed in the case that multiple eavesdroppers are randomly distributed in the VLC system with a practical amplitude constraint. The main contributions are summarized as follows.

1) The closed-form expressions for the upper and the lower bounds of the SOP and the ASC are derived by using the stochastic geometry method  and the derivation is very different from that in [5]. Moreover, it provides meaningful insight and guidance for the practical design of the VLC system with a amplitude constraint.

2) Accurate expressions with a low computational complexity for the SOP and the ASC are obtained and our derivations are more rigorous than that in [5]. Moreover, simulation results verify our theoretical analysis and show that the gap of the bounds on SOP is tighter than that in [11].

The rest of this letter is organized as follows. The system model and the preliminaries are presented in Section II. Section III analyzes the upper and the lower bounds of the SOP and the ASC. In Section IV, simulation results are presented. Finally, this letter is concluded in Section V.

\section{System Model and Preliminaries}
In this section, to facilitate the analysis, the system model is presented, and some preliminaries are clarified and derived.%In this section, the system model is presented. Moreover, to facilitate the analysis, some preliminaries are clarified and derived.
\subsection{System Model}
\begin{figure}[!t]
\setlength{\abovedisplayskip}{0pt}
\setlength{\belowdisplayskip}{0pt}
\centering
\includegraphics[width=1.5in]{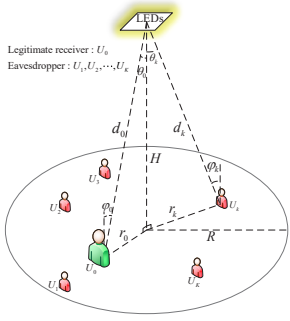}
\caption{The VLC system model.}
\label{system}
\end{figure}
As shown in Fig. 1, the downlink VLC system within a circular area of radius $R$ is considered. Similar to \cite{SOP_GAO}, $L$ LEDs are deployed in the center of the ceiling and closed to each other. Moreover, the LEDs provide illumination and communication services.Furthermore, the transmitted signal $\textbf{x}$ satisfies an amplitude constraint to avoid clipping distortion due to the limited dynamic range of the LED \cite{CC_AM}. Thus, the amplitude constraint is given as
\begin{align}
|\textbf{x}| \le \mathcal{A}.
\end{align}

Meanwhile, $K$ passive eavesdroppers exist in the VLC system and they can eavesdrop the confidential information. Without loss of generality, it is assumed that the legitimate receiver location is modeled as a uniform distribution and those  eavesdroppers are modeled as a homogeneous Poisson Point Process (HPPP) with an intensity $\lambda$ \cite{SOP_GAO}, \cite{3D_YL}.

According to \cite{SOP_GAO} and \cite{CC_AM}, the LoS path dominates the propagation and all the non-LoS paths can be neglected. Thus, the $i$th VLC channel gain is given as
{\setlength\abovedisplayskip{1pt}
\setlength\belowdisplayskip{1pt}
\begin{equation} \label{channel}
%h(z)=\frac{(m+1)A_k\cos\varphi_k}{2\pi d_k^2} \cos^m(\theta_k) g(\varphi_k) T_s(\varphi_k),
h_i=\frac{(m_{t,i}+1)A_{R,i}\cos\varphi_i}{2\pi {d_i}^2} \cos^m_{t,i}(\theta_i) g(\varphi_i) T_s(\varphi_i),
\end{equation}
where $A_{\!R,i\!}$ and ${\!d_i\!}$ are the detection area and the distance between the $i$th LED and the receiver, respectively. $\varphi_i$ and $\theta_i$ indicate the incident angle and the radiation angle shown in Fig. 1, respectively. $m_{t,i}\!=\!-{\rm log}(2)/{\!\rm log\!}(\!{\rm cos} \theta_{\!1/\!2,i}\!)$ denotes the order of Lambertian radiant with the $i$th LED semi-angle $\theta_{\!1\!/\!2,i}$. $T_s(\!\varphi_i\!)$ is the optical filter gain.  $g(\!\varphi_i\!)\!=\!n^2/{\rm sin^2} (\!\Psi_{{\rm FoV},i}\!)$ denotes the optical concentrator (OC) gain, which depends on the refractive index $n_i$ of the OC and the field of view $\Psi_{{\rm FoV},i}$ of the photodiode. Note that $h_i=0$ when $\varphi_i\!>\!\Psi_{{\rm FoV},i}$.}%semi-angle

\subsection{Preliminaries}

In order to utilize the statistical information of all receiver locations effectively, according to the work in \cite{SOP_GAO}, the relationship between the distance $d_{i,k}$ and the received RF power $P_{i,k}$ from the $i$th LED to the $k$th receiver can be given as
\begin{equation} \label{power_channel}
P_{i,k}=\mathcal{A}^2 (C_{i,k,RF}/d^4_{k}) G^2_{t,i} G^2_{r,i},
\end{equation}
where $C_{i,k,RF}$ is the RF power constant \cite{SOP_GAO}. $G^2_{t,i}\!=\!{\rm cos}^{m_{t,i,k}}\!(\!\theta_{i,k}\!)$ and $G^2_{r,i}\!=\!{\rm cos}^{m_{r,i,k}}\!(\!\varphi_{i,k}\!)$ are the radiation gain of the LED and the incidence gain of the receiver, respectively. Moreover,  $m_{t,i,k}\!=m_{t,i}$ and $m_{r,i,k}\!=\!-{\rm log}(2)/{\rm log}(\!{\rm cos} \Psi_{i,k}\!)$.%$m_{t,ik}$ and $m_{r,ik}$ are given by

It is assumed that all receivers (legitimate receiver and eavesdroppers) try to obtain the best performance for their own perspective. Thus, all photodiodes always face the radiation line \cite{SOP_GAO}. Based on the geometric relationship illustrated in Fig. 1, the received RF power in eq. (\ref{power_channel}) can be rewritten as
\begin{align} \label{power_simp}
P_{i,k}=\mathcal{A}^2 C_{i,k,RF} H^{2m_t} (H^2+r^2_{i,k})^{-2-m_{i,t}}.
\end{align}

For the VLC channels with the amplitude constraint and similar to the simplification process in \cite{SOP_GAO}, the peak signal-to-noise-ratio (SNR) for all receivers can be given as%{\color{blue}
\begin{equation} \label{SNR}
\gamma_{i,k} =L\mathcal{C}_{i,k} (H^2+r^2_{i,k})^{-2-m_{t,i}}, %=\frac{P_{k}}{N_0}
\end{equation}
where $\mathcal{C}_{i,k}\!=\!\mathcal{A}^2 C_{i,k,RF} H^{2m_{i,t}}\!/\!N_0$, and $N_0$ is the noise power.

Due to the amplitude constraint in eq. (1), the upper and the lower instantaneous secrecy capacity bounds of the VLC channel can be respectively given by \cite{CC_AM}.
\begin{subequations} \label{CC}
\begin{align}
&{\rm C_s^{upper}}=\max \left\{ \frac{1}{2}\log \left(\frac{\gamma_0 + 1}{\gamma^{\max}_E+1}\right), 0 \right\},\\
&{\rm C_s^{lower}}=\max \left\{ \frac{1}{2} \log \left(\frac{6\gamma_0 + 3\pi e}{\pi e \gamma^{\max}_E+3\pi e}\right),0 \right\},
\end{align}
\end{subequations}
where $\gamma^{\max}_E\!=\!\max \limits_{1\!\le \!k \le\! K}\left\{\!\gamma_{k}\!\right\}$ denotes the highest SNR among all eavesdroppers and $\gamma_0$ is the SNR of the legitimate receiver. %$e$ is the base of the natural logarithm.

For the HPPP model, one has $P(\!\Phi\!=\!k\!)\!=\![\mu^k {\rm exp}(\!-\mu\!)]/k!$, where $k\!=\!0,\!\ldots\!,K$ represents the $k$th eavesdropper; $\!\Phi$ denotes the number of eavesdropper, and $\mu\!=\!\lambda\pi R^2 $ is the mean of the HPPP. The probability density function (PDF) of the horizontal distance between the eavesdropper and the LED can be given as $f_r\!=\!\lambda/\mu\!=\!1/(\!\pi R^2\!)$, which is the same as the PDF of the legitimate receiver. Thus, the PDF of all receivers can be summarized as $f_r$. By using the stochastic geometry theory, the PDF  and the cumulative distribution function (CDF)  of $\gamma_k$ are respectively given by \cite{SOP_GAO}.%$f_{\gamma_{k}}(x)$ $F_{\gamma}(x)$
\begin{subequations}
\begin{align}
f_{\gamma_k}(x)&=ax^b, \\
F_{\gamma_k}(x)&=a x^c /c+\epsilon,
\end{align}
\end{subequations}
where $\epsilon\!=\!H^2\!/\!D^2\!+\!1$, $a\!=\!-\!c/\!(R^{2} \!\mathcal{C}^{\mathrm{c}})$, $b\!=\!c\!-\!1$, and $\mathrm{c}\!=\!-1 \!/\!(m_{t}\!+\!2)$. The variable $x \!\in \![\!\gamma_{\min}, \!\gamma_{\max}\!]$, and $\gamma_{\min}\!=\!\mathcal{C} (H^2\!+\!R^2)^{-(m_t\! +\!2)}$, $\gamma_{\max}\!=\!\mathcal{C} \!H^{-2(m_t+2)}$. Note that $\mathcal{C}\!\stackrel{\Delta}{=}\!\mathcal{C}_k$. By using the probability theory and the independent variables $\gamma_1, \dots, \gamma_K$, one has $F_{\gamma_E^{max}}(x)\!=\!\prod_{i\!=\!1}^{K}\! F_{\gamma_{k}}(x)$. Thus, the PDF of $\gamma_E^{max}$
 is given as
\begin{align}
f_{\gamma_{E}^{\max }}(x) &\!=\!a \!K \!\sum_{i=0}^{\!K\!-\!1} \mathrm{Q}\left(\frac{a}{c}\right)^{i} \!x^{c i+c-1} \!\epsilon^{\!K\!-1\!-\!i}\!=\!a \!K\! \sum_{i=0}^{\!K\!-\!1} \alpha_{i} x^{\beta_{i}},
\end{align}
where  $\small{\!\mathrm{Q}\!=\!\tbinom{\!K-\!1}{\!i}}$. Moreover, to simplify the expression, $\alpha_i$ and $\beta_i$ are used to replace $\mathrm{Q} {(\!\frac{a}{c}\!)}^i \epsilon^{\!K-\!1-\!i}$ and $\!ci+\!c-\!1$, respectively.
\section{SOP and ASC Analysis}
In this section, due to the similarity between the eq. (6a) and (6b), a general forms for the upper and the lower bounds of the SOP and the ASC are analyzed.%and lower bounds of the channel capacity
\subsection{ Upper and Lower Bounds of SOP}
Based on eq. (\ref{CC}), the upper and the lower bounds of the SOP can be respectively given as
\begin{subequations}\label{SOP_case}
\begin{align}
P_{\rm {SOP}}^{\rm {\! upper}} \label{SOP_upper_case}
&\!=\!\mathbb{P}(\!\gamma_{0}\! \leq\! \sigma_{\rm u}\! \gamma_{E}^{\!\max}\!+\!\zeta_{\rm u}\! )\! =\! \int_{y_{\!\min} }^{y_{\!\max}} \! \int_{x_{\! \min}}^{x_{\! \max}}\! f_{\gamma_{0}}(x)\! f_{\gamma^{\max }_E}(y)\! dx dy, \\
P_{\rm {SOP}}^{\rm {\! lower}} \label{SOP_lower_case}
&\!=\!\mathbb{P}(\!\gamma_{0}\! \leq\! \sigma_{\rm l}\! \gamma_{E}^{\!\max}\!+\!\zeta_{\rm l}\!)\!=\!\int_{y_{\!\min}}^{y_{\!\max} } \int_{x_{\!\min}}^{x_{\!\max}} \!f_{\gamma_{0}}(x)\! f_{\gamma^{\max }_E}(y)\! dx dy,
\end{align}
\end{subequations}
where $\sigma_{\rm u}=\frac{\pi e 2^{2C_{th}}}{6}$, $\sigma_{\rm l}=2^{2C_{th}}$ and $\zeta_{\rm u} =3\sigma_{\rm u}-\pi e/2$, $\zeta_l=\sigma_{\rm l}-1$. In the following analysis, the subscript $u$ and $l$ represent the upper and lower bound analysis, respectively. %$\sigma$ and $\zeta$ are replaced by $\sigma_u$ and $\zeta_u$, respectively, vice versa.  the bound analysis represents the upper bound when

Note that the integral interval should be discussed due to the uncertainty between $\gamma_{\max}$ and $\sigma y\!+\!\zeta$. Table I summarizes all the integral intervals, where $\gamma_{\rm limit}=\frac{\gamma_{\max}-\zeta}{\sigma}$.
\begin{table}[htbp]
\setlength{\abovedisplayskip}{-1.5cm}
\setlength{\belowdisplayskip}{-1.5cm}
\begin{center}
\caption{The integral interval of (\ref{SOP_case})}   \label{integral}
\begin{tabular}{|c|m{3.1cm}<{\centering}|m{0.6cm}<{\centering}|m{1.15cm}<{\centering}|m{0.6cm}<{\centering}|m{0.6cm}<{\centering}|} \hline
   \multicolumn{2}{|c|}{Case}  & $x_{\min}$  & $x_{\max}$  & $y_{\min}$  & $y_{\max}$ \\
  \hline
 1 &\tabincell{c}{$\sigma y+\zeta< \gamma_{\max}$\\ $ (\gamma_{\max}-\zeta)/\sigma > \gamma_{\min}$}  & $\gamma_{\min}$ & $\sigma y+\zeta$ & $\gamma_{\min}$ & $\gamma_{{\rm limit}}$  \\
  \hline
  2 &\tabincell{c}{$\sigma y+\zeta< \gamma_{\max}$\\ $(\gamma_{\max}-\zeta)/\sigma < \gamma_{\min}$} &$\gamma_{\min}$ & $\sigma y+\zeta$ & \multicolumn{2}{c|}{Empty set} \\
  \hline
 3 & \tabincell{c}{$\sigma y+\zeta> \gamma_{\max}$ \\ $ (\gamma_{\max}-\zeta)/\sigma > \gamma_{\min}$ }& $\gamma_{\min}$ & $\gamma_{\max}$ & $\gamma_{{\rm limit}}$ &$\gamma_{\max}$ \\
  \hline
  4 &\tabincell{c}{$\sigma y+\zeta> \gamma_{\max}$ \\$ \gamma_{\max}-\zeta/\sigma < \gamma_{\min}$}& $\gamma_{\min}$ & $\gamma_{\max}$ &  $\gamma_{\min}$ &  $\gamma_{\max}$\\
  \hline
\end{tabular}%}
\end{center}
\end{table}
It can be seen that cases 1, 3, and 4 need to be analyzed. The upper and the lower bounds of the SOP in case 1 can be given as
\begin{align} \label{SOP_case1} \notag
& P_{\rm {SOP1}}(\sigma, \zeta)=\int_{\gamma_{\min}}^{\gamma_{{\rm limit}}} \int_{\gamma_{\min}}^{\sigma y+ \zeta} f_{\gamma_{0}}(x) f_{\gamma^{\max}_E}(y) dx dy \\ \notag
&=\! \sigma^{\!b\!+\!1} \!\Theta \!\sum_{i\!=\!0}^{\!K\!-\!1}  \!\alpha_i \!\int_{\!\gamma_{\min}}^{\!\gamma_{{\rm limit}}} \!y^{\beta_i}\!\left(\!y\!+\!\frac{\zeta}{\sigma}\!\right)^{\!b\!+\!1} \!d\!y\!-\!\gamma_{\!\min}^{\!b\!+\!1}\!\Theta \!\sum_{i\!=\!0}^{\!K\!-\!1}\! \alpha_i \!\int_{\!\gamma_{\min}}^{\!\gamma_{{\rm limit}}} \!y^{\beta_i} d\!y \\ \notag
&=\!-\gamma_{\min}^{b+1}\Theta \sum_{i=0}^{K-1}  \alpha_i \left( \frac{\gamma_{{\rm limit}}^{\beta_i+1}}{\beta_i+1}-\frac{\gamma_{\min}^{\beta_i+1}}{\beta_i+1} \right) +\sigma^{b+1} \Theta \sum_{i=0}^{K-1} \alpha_i \\ \notag
&\times \left[ \frac{\delta^{b+1} {\gamma_{{\rm limit}}}^{\beta_i+1} {\rm H}\left([-b-1,\beta_i+1],\beta_i+2,-\frac{\gamma_{{\rm limit}}}{\delta} \right)}{\beta_i+1} \right. \\
&\left. -\frac{\delta^{b+1} {\gamma_{\min}}^{\beta_i+1} {\rm H}\left([-b-1,\beta_i+1],\beta_i+2,-\frac{\gamma_{\min}}{\delta} \right)}{\beta_i+1} \right],
\end{align}
where  $\delta\!=\!\frac{\zeta}{\sigma}$, $\Theta\!=\!\frac{a^2 K}{b+1}$ and ${\rm H}(\cdot)$ is the hygergeom function.

Since the unique difference between  case 3 and 4 is the integral interval of the variable $y$, in order to simplify this derivation process, the upper and the lower bounds of the SOP in case 3 and 4 are derived together as
\begin{align} \label{SOP_case34} \notag
P_{\rm {SOP3,4}}&(\sigma, \zeta) =\int_{y_{\min}}^{\gamma_{\max}} \int_{\gamma_{\min}}^{\gamma_{\max}} f_{\gamma_{0}}(x) f_{\gamma^{\max}_E}(y) dx dy \\ \notag
&= a^2 K \sum_{i=0}^{K-1}  \alpha_i \int_{y_{\min}}^{\gamma_{\max}} y^{\beta_i} \int_{\gamma_{\min}}^{\gamma_{\max}} x^{b} dx dy \\
%&= \Theta \sum_{i=0}^{K-1} \alpha_i \left(\gamma_{\max}^{b+1}-\gamma_{\min}^{b+1}\right)\int_{y_{\min}}^{\gamma_{\max}} y^{\beta_i} dy\\
%&  \qquad -\gamma_{\min}^{b+1}\Theta \sum_{i=1}^{K-1}  \alpha_i \int_{\gamma_{\min}}^{\gamma_{{\rm limit}}} y^{\beta_i} dy \\ \notag
&= \Theta \sum_{i=0}^{K-1} \alpha_i \left(\gamma_{\max}^{b+1}-\gamma_{\min}^{b+1}\right) \frac{\gamma_{\max}^{\beta_i+1}-y_{\min}^{\beta_i+1} }{\beta_i+1},
\end{align}
where $y_{\min}$ is case dependent (case 3 or 4).

Thus, the  upper and the lower bounds of the SOP can be respectively summarized as
\begin{subequations} \label{SOP}
\begin{align}
&P_{\rm SOP}^{\rm upper}= P_{\rm {SOP1}}(\sigma_u, \zeta_u)+P_{\rm {SOP3,4}}(\sigma_u, \zeta_u),\\
&P_{\rm SOP}^{\rm lower}= P_{\rm {SOP1}}(\sigma_l, \zeta_l)+P_{\rm {SOP3,4}}(\sigma_l, \zeta_l).
\end{align}
\end{subequations}
\subsection{Upper and Lower Bounds of ASC}
The ASC is the expected value of the instantaneous secrecy capacity ${\rm C_s}$. The upper and the lower bounds of the ASC can be summarized as
\begin{align}  \notag
&{\rm \overline C_s^{u, l}} = \int_{\gamma_{\min}}^{\gamma_{\max}}\int_{\gamma_{\min}}^{\gamma_{\max}}{\rm{C_s^{u,l}}} f (\gamma_0, \gamma_E^{\max}) d\gamma_0 d\gamma_E^{\max} \\ \notag
&=\frac{1}{2\ln2} \bigg[\underbrace { \int_{\gamma_{\min}}^{\gamma_{\max}} \ln(m_0 x+n_0) f_{\gamma_0}(x)\int_{\gamma_{\min}}^{x}f_{\gamma_E^{\max}}(y) dydx}_{\rm \overline C_{s1}^{u,l}} \\
&-\underbrace{\!\int_{\gamma_{\min}}^{\gamma_{\max}} \ln(\!m_E y+n_E\!) f_{\gamma_E^{\max}}(y)\int_{y}^{\gamma_{\max}}f_{\gamma_0}(x) dxdy }_{\rm \overline C_{s2}^{u,l}}\!\bigg],
\end{align}
where $f(\!\gamma_0, \!\gamma_E^{\max}\!)$ denotes the joint PDF of the legitimate receiver and eavesdroppers. Moreover, all channels are independent. The  upper bound of the ASC is obtained when ${\rm{C_s^{u,l}}}\!=\!{\rm{C_s^{upper}}}$, $m_0\!=\!1$, $n_0\!=\!1$, $m_E\!=\!1$, and $n_E\!=\!1$. The lower bound of the ASC is obtained when ${\rm{C_s^{u,l}}}\!=\!{\rm{C_s^{lower}}}$, $m_0\!=\!6$, $n_0\!=\!3\pi e$, $m_E\!=\!\pi e$, and $n_E\!=\!3\pi e$. In order to simplify the analysis, a function is defined as
\begin{align} \notag
&{\rm LP}(\tau, \!p,\! q)\!=\!\int_p^q \!x^{\tau} \!\ln \!(1\!+\!x\!)\!dx = \frac{q^{\tau+1} \!\ln\! (\!q\!+\!1\!)}{\tau+1} -\frac{p^{\tau+1} \!\ln \!(\!p\!+\!1\!)}{\tau+1}\\ \notag
&+\frac{q^{\tau}[\tau^2(1-q)+\tau(3-2q)+2]}{\tau(\tau+2)(\tau+1)^2} +  \frac{q^{\tau}(-2-\tau) \Phi (-q,1,\tau)}{(\tau+1)(\tau+2)}\\
&-\!\frac{p^{\tau}\![\tau^2\!(\!1\!-p\!)\!+\tau\!(\!3\!-\!2p\!)+\!2]}{\tau(\tau+2)\!(\tau+1)^2}\! - \! \frac{p^{\tau}(\!-2\!-\tau) \Phi (\!-p,\!1,\!\tau)}{(\tau+1)(\tau+2)},
\end{align}
%-\frac{p^{\tau}(-\tau^2 p+\tau^2-2\tau p+3\tau+2)}{\tau(\tau+2)(\tau+1)^2}
where $\Phi(\cdot)$ is the LerchPhi function \cite{math} and can be calculated by Maple software.

Subsequently $\rm \overline C_{s1}^{u,l}$ and $\rm \overline C_{s2}^{u,l}$ can be given as
\begin{subequations}
\begin{align} \label{Cs1} \notag
&{\rm \overline C_{s1}^{u,l}}= \int_{\frac{m_E\gamma_{\min}}{m_0}}^{\gamma_{\max}} a x^b \ln(\!m_0 \!x+\!n_0\!) \int_{\gamma_{\min}}^{\frac{m_0x}{m_E}} aK \sum_{i=0}^{K-1} \alpha_i y^{\beta_i} dydx\\ \notag
&=\Delta_1 \! \bigg[\!\ln\!(\!n_0\!)\big(\!\frac{(\gamma_{\max}^{\kappa_i}\!-{\!\frac{m_E\!\gamma_{\min}}{m_0}}^{\kappa_i})}{\kappa_i(\!m_0/m_E\!)^{-(\beta+1)}}-\!\frac{\!\gamma_{\min}^{\!\beta_i+1}(\gamma_{\max}^{b+1}-\!{\frac{m_E\!\gamma_{\min}}{m_0}}^{b+1})}{b+1}\big) \\ \notag
&\quad +(\frac{m_0}{m_E})^{\beta+1}(\frac{n_0}{m_0})^{\beta_i+b+2}{\rm LP}(\kappa_i-1,\frac{m_E{\gamma_{\min}}}{n_0},\frac{m_0{\gamma_{\max}}}{n_0}) \\
&\quad -\gamma_{\min}^{\beta_i+1}(\frac{n_0}{m_0})^{b+1}{\rm LP}(b,\frac{m_E{\gamma_{\min}}}{n_0},\frac{m_0{\gamma_{\max}}}{n_0}) \bigg], \\ \notag
&{\rm \overline C_{s2}^{u,l}}= \!\int_{\gamma_{\min}}^{\frac{m_0\gamma_{\max}}{m_E}} \!aK\ln(\!m_E y\!+\!n_E\!) \sum_{i=0}^{K-1} \alpha_i y^{\beta_i}  \int_{\frac{m_Ey}{m_0}}^{\gamma_{\max}} a x^b dxdy\\ \notag
&\!=\Delta_2 \! \bigg[\!\ln\!(\!n_E\!)\big(\!\frac{\gamma_{\max}^{b+1}\!(\!{\frac{m_0\!\gamma_{\max}}{m_E}}^{\!\beta_i\!+1}\!-\!\gamma_{\min}^{\beta_i+1})}{\beta_i+1}-\!\frac{(\!{\frac{m_0\!\gamma_{\max}}{m_E}}^{\kappa_i}-\!\gamma_{\min}^{\kappa_i})}{\kappa_i(\!m_E/m_0\!)^{-(b+1)}} \big)\\ \notag
&-(\frac{m_E}{m_0})^{(b+1)}(\frac{n_E}{m_E})^{\kappa_i}{\rm LP}(\kappa_i-1,\frac{m_E}{n_E}{\gamma_{\min}},\frac{m_0}{n_E}{\gamma_{\max}}) \\
&+\gamma_{\max}^{b+1}(\frac{n_E}{m_E})^{\beta_i+1}{\rm LP}(\beta_i,\frac{m_E}{n_E}{\gamma_{\min}},\frac{m_0}{n_E}{\gamma_{\max}}) \bigg],
\end{align}
\end{subequations}
where $\Delta_1\!=\!\sum_{i=0}^{K\!-\!1}\! \frac{a^2\!K \alpha_i}{\beta_i+1}$, $\Delta_2\!=\!\sum_{i=0}^{\!K\!-\!1} \!\frac{a^2K \alpha_i}{b+1}$, $\kappa_i\!=\!\beta_i\!+\!b\!+2$.

\section{Simulation Results}
In this section, simulation results based on the Monte Carlo are presented to demonstrate the correctness of the theoretical analysis on the SOP and the ASC. The parameters are selected based on the work in \cite{SOP_GAO}, given as: $L=4$, $\theta_{1/2}\!=\!70^\circ$, $H\!=\!2.5$ m, $C_{th}\!=\!1 $ bit/s/Hz, $N_{0}\!=\!-98.82$ dBm, and $\mathcal{A}\!=\!6$ V.

It can be seen from Fig. 2 and Fig. 3 that the simulation results of the upper and the lower bounds of the SOP and the ASC match well with the theoretical results, which verify the accuracy of our theoretical analysis. For the SOP performance, the gap between the upper and the lower bounds is tight. Moreover, the SOP performance deteriorates with the increase of the number of eavesdroppers that result from the increase of  either $\lambda$, or $R$, or both. For the ASC performance, it can be seen from Fig. 3 that the ASC deteriorates with the increase of $\lambda$ or $R$, since the probability of eavesdroppers being closer to the LEDs is gradually higher than that of the legitimate receiver being closer to the LEDs. Moreover, the constraint that $K \geq 1$ in [5] is relaxed to obtain a reasonable setting since the probability that the number of the eavesdroppers equals to zero is dominant in the case of small intensity $\lambda$ with the HPPP model. Meantime, the ASC performance has improved significantly.
\section{Conclusion}
\begin{figure}[!t]
\centering
\includegraphics[width=2.4in]{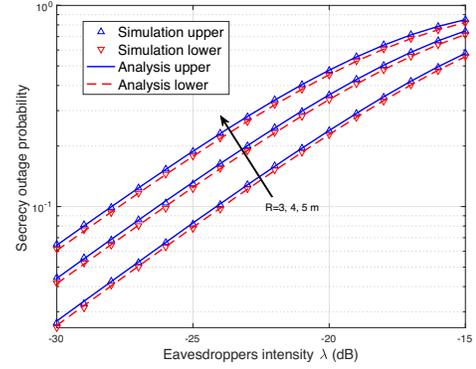}
\caption{SOP versus the eavesdropper intensity $\lambda$.}
\label{SOP_result}
\end{figure}

\begin{figure}[!t]
\centering
\includegraphics[width=2.4in]{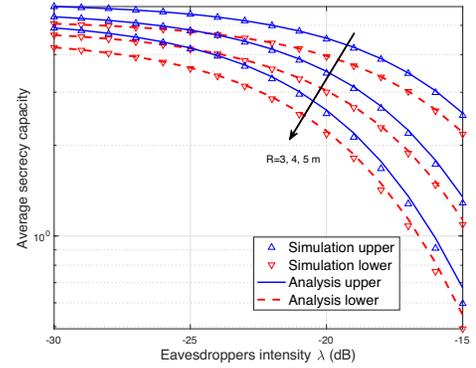}
\caption{ASC versus the eavesdropper intensity $\lambda$.}
\label{ASC_result}
\end{figure}
The security performance analysis for the VLC system with random location of eavesdroppers under the practical amplitude constraint was  studied. For the case that the locations of the legitimate receiver and eavesdroppers are unknown, the closed-form expressions for the upper and the lower bounds of the SOP and the ASC were derived by using the stochastic geometry method. Simulation results verified the correctness of the theoretical analysis. Moreover, the analysis method proposed in this paper can be easily extended into the case with the collaborative multiple eavesdroppers.

\end{document}